\documentclass[11pt]{article}

\usepackage{amsmath}
\usepackage{amssymb}

\def\a{\alpha}
\def\b{\beta}

\def\m{\mu}
\def\n{\nu}

\def\w{\omega}
\def\ad{{{\dot{\alpha}}}}
\def\bd{{{\dot{\beta}}}}
\def\gd{{{\dot{\gamma}}}}

\makeatletter
\@addtoreset{equation}{section}

\makeatother

\newcommand{\tr}{\mathop{\mathrm{tr}}}
\newcommand{\Pf}{\mathop{\mathrm{Pf}}}

\newcommand{\deriv}[2]{\dfrac{\partial #1}{\partial #2}}

\newcommand{\cN}{{\mathcal N}}
\newcommand{\cO}{{\mathcal O}}
\newcommand{\cC}{{\mathcal C}}
\newcommand{\cQ}{{\mathcal Q}}

\newcommand{\bZ}{{\mathbb Z}}
\newcommand{\bC}{{\mathbb C}}
\newcommand{\rT}{{\mathrm T}}
\newcommand{\rg}{{\mathrm g}}
\newcommand{\tQ}{{\widetilde Q}}
\newcommand{\tp}{{\widetilde\psi}}
\newcommand{\tJ}{{\widetilde J}}
\newcommand{\tM}{{\widetilde M}}

\begin{document}

\begin{titlepage}

\vspace*{-15mm}   
\baselineskip 10pt   
\begin{flushright}
\begin{tabular}{l}
{March 2008} \\
{APCTP Pre2008-001}\\
{SU-ITP-08/05}\\
{arXiv:yymm.nnnn}
\end{tabular}
\end{flushright}
\baselineskip 24pt   
\vglue 10mm   

\renewcommand{\thefootnote}{\fnsymbol{footnote}}

\vspace{15mm}
\baselineskip 9mm
\begin{center}
{\LARGE \bf D-instanton derivation of multi-fermion $F$-terms 
\\
in supersymmetric QCD
}

\end{center}

\baselineskip 6mm
\vspace{10mm}
\begin{center}
Yoshinori Matsuo,$^a$%
\footnote{\tt ymatsuo@apctp.org}
Jaemo Park,$^{b,c,d}$%
\footnote{\tt jaemo@postech.ac.kr} 
Cheol Ryou,$^b$%
\footnote{\tt string@postech.ac.kr}
and 
Masayoshi Yamamoto$^a$%
\footnote{\tt yamamoto@apctp.org} 
\\[5mm] 
{\it $^a$Asia Pacific Center for Theoretical Physics, POSTECH, 
 Pohang 790-784, Korea \\ 
 $^b$Department of Physics, POSTECH, Pohang 790-784, Korea \\
 $^c$Postech Center for Theoretical Physics (PCTP), Pohang, 790-784,
 Korea \\
 $^d$Department of Physics, Stanford University, Stanford, CA 94305-4060, USA}
\end{center}

\thispagestyle{empty}

\vfill
\begin{abstract}
We investigate effects of field theory instantons 
by considering D-instantons in a suitable D3-brane background. 
In supersymmetric QCD with $SU(N_c)$ gauge group 
with $N_f=N_c$ flavors, 
the moduli space of vacua is deformed by instantons. 
This effect can be described by 
the chiral interactions which are called 
multi-fermion $F$-terms. 
We derive these chiral interaction terms 
as D-instanton effects in the presence of D3-branes. 
For $SU(2)$, the obtained result agrees with the previous
result worked out by Beasley and Witten \cite{Beasley:2004ys}.
We also explicitly work out 
those for the case of the symplectic gauge group, 
and show that they describe the deformation of the moduli space. 
\end{abstract}
 
\vspace{20mm}
\end{titlepage}
\setcounter{footnote}{0}
\renewcommand{\thefootnote}{\arabic{footnote}}

\vspace{1cm}

\section{Introduction}\label{sec:Intro}

Instanton physics in string theory is an interesting area to study.
There are various nonperturbative objects such as D-branes, membranes 
and 5-branes and one can consider the instanton effects arising from 
such objects wrapping on some suitable cycles 
\cite{Witten:1995gx, Douglas:1995bn, Witten:1996bn}. 
Spacetime approach or the physical gauge approach was initiated by 
\cite{Harvey:1999as} and various interesting works were done. 
Since some field theories can be embedded in a string theory, field
theory instantons can be understood in terms of instantons in string
theory and this approach shed much light on the understanding of the 
structures of gauge theory instantons, such problem as the measure of
the instanton moduli space. One of the important discovery
is the D5-brane as a small instanton in Type I theory \cite{Witten:1995gx} 
and this led to the much progress in our understanding 
of the field theory instantons. 
Also in the string theory set up, one can have truly stringy
instantons in the embedded field theory 
\cite{Blumenhagen:2006xt, Ibanez:2006da, 
Florea:2006si, Abel:2006yk, Argurio:2007vqa, Aharony:2007pr}, 
whose effect cannot be reproduced within
field theory and this leads to many interesting physics such as
dynamical supersymmetry breaking combined with the other effects. 

In this work we will use the string theory setup to understand one
aspect of gauge theory instanton \cite{Billo:2002hm}. 
In particular we are interested in 
$\cN=1$ supersymmetric QCD and derive one interesting effect coming
from the gauge theory instanton. We realize the gauge theory as a
suitable D3-brane configuration \cite{Bertolini:2001gg} 
where  D-instanton plays the role
of the usual gauge theory instanton.
In $\cN=1$ $SU(N_c)$ gauge theory with $N_f$ fundamental flavors, the well known 
instanton effect is the generation of 
the ADS superpotential \cite{Affleck:1983mk} for
$N_c=N_f-1$, which lifts the moduli of vacua 
(for a review, see \cite{Intriligator:1995au}).
This superpotential can be reproduced by 
the D-brane effective theory \cite{Akerblom:2006hx}. 
For $N_f=N_c$, no superpotential is generated but a quantum effect 
deforms the complex structure of the moduli space \cite{Seiberg:1994bz}. 
In \cite{Beasley:2004ys}, it was pointed out 
that the deformation of the moduli space 
is still due to the usual gauge theory instanton, 
which gives rise to a chiral interaction. 
This interaction is a four-fermion interaction term 
and called multi-fermion $F$-terms. 
For $N_f>N_c$, instantons generate 
interaction terms with more fermions. 

In this paper, we reproduce 
these multi-fermion $F$-terms 
from the D-brane effective theory, which is realized as D3-branes 
at a particular orbifold singularity. 
Though only the scalars which parametrize 
the moduli of $\cN=1$ SQCD are involved 
in the calculation of the ADS superpotential, 
we have to include interactions 
of fermions in the quark superfields. 
These $F$-terms are given 
in the form of the integral with respect to 
the moduli of the instanton. 
It is difficult to perform 
the integration for general case. 
Hence, we calculate 
the simplest case $N_c=N_f=2$. 
In this case, we can easily perform 
the integration and obtain the 
multi-fermion $F$-terms which 
are equivalent to the result of \cite{Beasley:2004ys}. 
We study the case of 
the symplectic group as another example, 
and make a connection with 
the deformation of the moduli space. 
In this case, we can also explicitly carry out the  integration,
since the ADHM constraint is absent and 
the structure of the instanton moduli is simple. 

This paper is organized as follows. 
In section \ref{sec:Prelim}, we briefly recall 
the basic setup of the D-brane effective theory 
which describes instantons 
in the $\cN=4$ super Yang-Mills \cite{Billo:2002hm}. 
In section \ref{sec:Orbi}, we introduce D3-branes on one particular orbifold 
and realize the $\cN=1$ SQCD with $SU(N_c)$ gauge group \cite{Bertolini:2001gg}. 
After that, we discuss basic facts 
about the derivation of superpotentials 
and show the general expression 
of multi-fermion $F$-terms for $N_f=N_c$. 
In section \ref{sec:Rev}, we briefly review the 
multi-fermion $F$-terms and 
their relation to the deformation of the moduli space \cite{Beasley:2004ys}. 
In section \ref{sec:SU(2)}, 
we describe details of calculations 
in the simplest case of $N_f=N_c=2$. 
We show that our result coincides
with that in \cite{Beasley:2004ys}. 
In section \ref{sec:USp}, 
we consider the more general case of $USp(N_c)$. 
We derive an expression of multi-fermion $F$-terms 
and show that they describe the deformation of 
the moduli space for $N_f=N_c+1$. 
We also consider multi-fermion $F$-terms 
with more fermions, which appear for $N_f>N_c+1$. 
Section \ref{sec:Concl} is devoted for conclusions and discussions.

\section{Preliminaries}\label{sec:Prelim}

In this section, 
we explain the basic setup of the system. 
It is well known that 
the D-instanton is the gauge theory instanton if the gauge theory
in consideration is realized by D3-brane configurations \cite{Billo:2002hm}.
Here we describe the massless spectrum of various sectors coming 
from D3 branes and D-instantons in the flat space 
and specify the interaction terms 
between various sectors. 
We divide open string fields into the following 
three sectors. 

\paragraph{D3-D3 sector}
This sector consists of the open strings 
whose both ends are on the D3-branes. 
At low energy, only massless modes of 
these open strings contribute to the theory. 
These massless fields form $\cN=4$ super Yang-Mills multiplets.
Let $x^{\mu}$ denote the world-volume coordinates of D3-branes and 
$x^a$ denote the transverse coordinates of the remaining six dimensions. 
The bosonic components are denoted by $A^{\mu}$ and $X^a$, and 
their fermionic partners by $\Psi_{\alpha}^A$ and
$\bar{\Psi}^{\dot{\alpha}}_A$. 
Indices of $A$ are those of $SU(4)$ denoting the chirality of 
the transverse six dimensions while $\alpha$ and $\dot{\alpha}$ 
denote the usual four dimensional chirality. 
We also write six scalars $X^a$ in the
antisymmetric representation 
\begin{equation}
\Phi_{AB}\equiv \bar \Sigma^a_{AB}X^a.
\end{equation}
where $\bar{\Sigma}^a$ (and $\Sigma^a$) realize six-dimensional 
Clifford algebras, and appear in the expression of six-dimensional
$\gamma$-matrices, 
\begin{equation}
\Gamma^a= \left( \begin{array}{cc}
                  0 & \Sigma^a\\
                  \bar{\Sigma}^a & 0 \end{array}  \right).
\end{equation}
When we put $N$ D3-branes, 
all of theses fields are in the adjoint representation of $SU(N)$, 
and can be written as $N \times N$ hermitian matrices. 

\paragraph{D(-1)-D(-1) sector}

This sector consists of the open strings with 
both ends on the D-instantons. 
These are obtained by the dimensional reduction of 
ten-dimensional super Yang-Mills theory. 
The bosonic fields are written as $a^{\mu}, \chi^a$ 
and the fermionic modes are denoted as 
$M^{\alpha A}, \lambda_{\dot{\alpha} A}$. 
Here  we have adopted the ADHM inspired notation. 
We also introduce the triplet of the auxiliary fields $D^c$. 
These are expressed in terms of $k\times k$ matrices 
for a background with $k$ D-instantons.
It is argued in \cite{Billo:2002hm} that 
there are subtleties to obtain the D-instanton action
in the presence of D3-branes by taking $\alpha'\to 0$ limit.
This is because the coupling constant dependence on $\alpha'$ is
different for D3-branes and D-instantons. If we take the coupling constant of
D3-branes to be constant, the coupling constant of D-instanton
diverges. In order to obtain the usual interacting theory for D(-1) massless
modes, we need a suitable rescaling of the fields on D-instantons.
Here we assume that such rescaling is already taken.
We will refer to the rescaled fields 
by using the above notations. 

\paragraph{D3-D(-1) sector}

This sector includes the massless modes 
of the open strings stretching 
between the D3-branes and D-instantons. 
From the Neveu-Schwarz(NS) sector we have a bosonic spinor in the first
four-directions 
where the GSO projection picks up the negative chirality. 
In the conjugate sector
we obtain an independent bosonic spinor with 
the same chirality. 
We will refer to them as $\omega_\ad$ 
and $\bar \omega_\ad$, respectively. 
From the Ramond sector and its conjugate sector, 
we obtain a pair of fermions 
$\mu^A$ and $\bar{\m}^A$. 
These fields are $N\times k$ and 
$k\times N$ matrices, respectively. 
\\

We are specifying the action of various sectors
and the SUSY transformations
are worked out in appendix \ref{SUSY}. 
Massless modes of the D(-1)-D(-1) and D3-D(-1) sectors 
correspond to the moduli of the gauge instanton 
which appear in the ADHM construction 
\cite{Atiyah:1978ri, Dorey:2002ik}. 
When carrying out the path integral, one obtains the measure of 
the instanton moduli space naturally. 
The action is given by
\begin{align}
S_1&= \tr\biggl(-[\a_{\m},
  \chi^a]^2+\chi^a\bar{\w}_{\dot{\a}}\w^{\dot{\a}}\chi^a
+\frac{1}{2}(\bar{\Sigma^a})_{AB}\bar{\m}^A\m^B\chi_a
-\frac{i}{4}(\bar{\Sigma^a})_{AB}M^{\a A}[\chi_a, M^B_{\a}] \notag\\
 &\quad +i(\bar{\m}^A\w_{\dot{\a}}+\bar{\w}_{\dot{\a}}\m^A
+\sigma^{\m}_{\b\dot{\a}}[M^{\b A}, a_{\m}])\lambda_A^{\dot{\a}}
-iD^c (\bar{\w}^{\dot{\a}}(\tau^c)^{\dot{\b}}_{\dot{\a}}\w_{\dot{\b}}
+i\bar{\eta^c}_{\m\n}[a^{\m}, a^{\n}])\biggr)
\end{align}
where $\eta, \bar{\eta}$ are the usual 't Hooft symbols and $\tau^c$ 
denote the usual Pauli matrices. 

Including interaction terms with the scalars in 
$\cN=4$ super Yang-Mills theory, 
we can also obtain the instanton effective action 
with non-zero VEVs of the scalars, which is given by, 
\begin{align}
 S_2 =\tr \left(
 \frac{1}{8}\varepsilon^{ABCD}\bar{\w}_{\ad}\Phi_{AB}\Phi_{CD}\w^{\ad}
 +\frac{i}{2}\bar{\m}^A\Phi_{AB}\m^B 
 +\frac{1}{4}\varepsilon^{ABCD}\bar{\w}_{\ad}\chi_{AB}\Phi_{CD}\w^{\ad}\right) . 
\end{align}

In this paper, we also include contributions 
from the fermions in the D3-brane field theory. 
The interaction terms with the fermions (and gauge fields) 
are as follows: 
\begin{align}
S_3&=\tr\left(i\bar{\w}_{\ad}\bar{\Psi}^{\ad}_A\m^A
 -i\bar{\m}^A\bar\Psi_{\ad A}\w^{\ad}
 +\frac{1}{2}\bar{\w}_{\ad}\bar{\sigma}^{\m\n\ad}_{\bd}\w^{\bd}F_{\m\n} \right). 
\end{align}
We will mainly consider nonzero fermionic background
with no gauge field background.

\section{Multi-fermion $F$-terms from D-instantons}\label{sec:Orbi}

In this paper, we study 
nonperturbative effects of instantons 
in an $\cN=1$ gauge theory. One way to obtain $\cN=1$ theory
is the orbifolding with putting D3-branes at 
the orbifold singularity \cite{Bertolini:2001gg}. 
Here we take a $\bC^3 / \bZ_2\times\bZ_2$ orbifold. 
We also introduce fractional D3-branes 
which are D5-branes wrapped on the singularity. 
Then, the gauge group is broken into 
$U(N_1)\times U(N_2)\times U(N_3)\times U(N_4)$. 
We assume that some of $U(1)$ factors can get massive, e.g., 
via Green-Schwarz mechanism
and we can get $SU(N)$ factors. When we discuss $SU(N)$ theory, we
assume that $U(1)$ is decoupled.  
We will refer to D3-brane configurations as $(N_1,N_2,N_3,N_4)$. 
The configuration of $(N_c,N_f,0,0)$ 
yields the $\cN=1$ SQCD with gauged flavor symmetry. 
We neglect gauging of 
flavor symmetry and 
treat the model as the usual $\cN=1$ SQCD. 
Configurations of D-instantons (and fractional D-instantons) 
are characterized by a similar fashion, $(k_1,k_2,k_3,k_4)$. 
It is taken to be $(1,0,0,0)$ in order to 
describe the one-instanton background. 

In order to obtain the $\bC^3 / \bZ_2\times\bZ_2$ orbifold,
we introduce the following complex coordinates, 
\begin{align}
 z^1 &= x^4 + i x^5 , & 
 z^2 &= x^6 + i x^7 , & 
 z^3 &= x^8 + i x^9.
\end{align}
The orbifold is constructed from the following 
two projections 
\begin{subequations}\label{Zorbifold}
 \begin{align}
  g_1 &: \qquad
  z^2 = -z^2 \quad\text{and}\quad 
  z^3 = -z^3 , \\ 
  g_2 &: \qquad
  z^1 = -z^1 \quad\text{and}\quad 
  z^3 = -z^3 . 
 \end{align}
\end{subequations}

The low energy effective theory of the D3-brane 
becomes the $\cN = 1$ quiver gauge theory with 
bi-fundamental matters. 
This theory can be obtained by imposing an orbifold projection 
on the $\cN = 4$ SYM. 
Since the orbifolding is abelian, 
the orbifold projection act on the Chan-Paton factor 
as follows: 
\begin{align}
 A_\mu & = \gamma(g_i) A_\mu \gamma(g_i)^{-1} , & 
 \Phi_{AB} &= \pm \gamma(g_i) \Phi_{AB} \gamma(g_i)^{-1} ,  \,\, i=1,2
\end{align}
where the $\pm$ sign must be chosen according to 
the projections \eqref{Zorbifold}. 
Since the orbifold is abelian, 
the representation matrix $\gamma(g_i)$ 
can be diagonalized and written in terms of 
the block diagonal matrices:  
\begin{align}
 \gamma(g_1) &= \left(
 \begin{matrix}
  1 & 0 & 0 & 0 \\
  0 & 1 & 0 & 0 \\
  0 & 0 & - 1 & 0 \\
  0 & 0 & 0 & - 1 
 \end{matrix}
 \right) \ ,  & 
 \gamma(g_2) &= \left(
 \begin{matrix}
  1 & 0 & 0 & 0 \\
  0 & - 1 & 0 & 0 \\
  0 & 0 & 1 & 0 \\
  0 & 0 & 0 & - 1 
 \end{matrix}
 \right) \ , 
\end{align}
where size of each blocks is determined by 
the D3-brane and D-instanton configurations. 
They have the size of $(N_1,N_2,N_3,N_4)$ 
for the Chan-Paton factors on D3-branes, 
and $(k_1,k_2,k_3,k_4)$ for those on D-instantons. 

Now, we take the configuration of $\cN=1$ SQCD, 
namely, $(N_c,N_f,0,0)$ for D3-branes and 
$(1,0,0,0)$ for D-instantons. 
The fields which survive in the orbifolding are as follows. 
First, we consider fields on the D3-branes. We have the 
gauge group $SU(N_c) \times SU(N_f)$ and regard the former
as a color gauge group and the latter as a flavor group.
Among the scalar fields, only two real (one complex) scalars 
have non-zero components. 
In terms of $SU(4)$ R-symmetry, 
$\Phi_{23}$ and $\Phi_{14}$ survive, 
and they are hermitian conjugate to each other. 
Non-zero components are 
\begin{align}
 \Phi_{23} &= \left(
              \begin{matrix}
               0 & Q & 0 & 0 \\
               \widetilde Q & 0 & 0 & 0 \\
               0 & 0 & 0 & 0 \\
               0 & 0 & 0 & 0
              \end{matrix}
             \right) , & 
 \Phi_{14} &= \left(
 \begin{matrix}
  0 & \tQ^\dagger & 0 & 0 \\
  Q^\dagger & 0 & 0 & 0 \\
  0 & 0 & 0 & 0 \\
  0 & 0 & 0 & 0
 \end{matrix}
 \right) . 
\end{align}
These two scalars are bi-fundamental in $(N_c, N_f)$.
Concerning fermions, two fermions $\Psi^1$ and $\Psi^4$, 
(and their hermitian conjugate) survive. 
The fermion $\Psi^1$ has a non-zero component in a similar fashion 
to the scalar fields: 
\begin{align}
 \Psi^1 &= \left(
           \begin{matrix}
            0 & \psi & 0 & 0 \\
            \widetilde \psi & 0 & 0 & 0 \\
            0 & 0 & 0 & 0 \\
            0 & 0 & 0 & 0
           \end{matrix}\right) , & 
 \bar \Psi_1 &= \left(
           \begin{matrix}
            0 & \bar\tp & 0 & 0 \\
            \bar \psi & 0 & 0 & 0 \\
            0 & 0 & 0 & 0 \\
            0 & 0 & 0 & 0
           \end{matrix}\right) . 
\end{align}
Fermions $\psi$ and $\tp$ are 
the superpartner of the scalars $Q$ and $\widetilde Q$ 
respectively. 
The other surviving fermion $\Psi^4$ becomes the gaugino $\lambda^{(\rg)}$
and has two non-zero block-diagonal components. 

For the fields in the D(-1)-D(-1) sector, 
the situation is similar, 
but much more components vanish. 
Since we set all block to be zero except for the first, 
only the first diagonal block survives. 
Furthermore, $a_\mu$ and $M^{\alpha A}$ do not 
appear in the action. 
Then, relevant fields in this sector are 
$\lambda^\ad\equiv\lambda_4^\ad$ and $D^c$. 
In the D3-D(-1) sector, the first diagonal components of 
$\omega_\ad$, $\bar\omega_\ad$, $\mu\equiv\mu^4$, $\bar\mu\equiv\bar\mu^4$ 
survive, and are in the fundamental and anti-fundamental 
representations of $SU(N_c)$, respectively. 
Off-diagonal components of $\mu'\equiv \mu^1$ and $\bar\mu'\equiv \bar\mu^1$ 
also survive and they are in the fundamental and anti-fundamental 
representations of $SU(N_f)$, respectively. 

After the $\bZ_2 \times \bZ_2$ orbifold projection, 
the action is obtained from the action of D3 and D(-1) branes
after keeping terms compatible with the orbifold projections.
The part containing only the fields in D(-1)-D(-1) and D3-D(-1) 
sectors is greatly simplified and 
turns out to be 
\begin{equation}
S_1=i(\bar{\m}_u\w^u_{\ad}+\bar{\w}_{\ad
  u}\m^u)\lambda^{\ad}-iD_c\bar{\w}^{\ad}_u(\tau^c)_{\ad}^{\,\,
  \bd}\w^u_{\bd} . 
\end{equation}
The interaction terms with scalars are 
\begin{equation}
S_2=\frac{1}{2}\bar{\w}_{\ad \, u}(Q^u_f{Q}^{\dagger f}_v
+\tQ^{\dagger u}_f\tQ^f_v)\w^{\ad
  v}-\frac{i}{2}\bar{\m}_u\tQ^{\dagger \,
  u}_f\m^{' f}+\frac{i}{2}\bar{\m}^{'}_f Q^{\dagger \, f}_u\m^v . 
\end{equation}
The above action is equivalent 
to those appearing in the ADHM construction. 
Here, we also include the interaction terms 
with fermions. 
These terms and the interactions with gauge fields and gauginos are 
\begin{align}
S_3&= -i\bar{\m}^{'}_f\bar{\psi}^f_{\ad u}\w^{\ad u}
 +i\bar{\w}_{\ad \, u}\bar{\tp}^{\ad \, u}_f{\m'}^f
 -i\bar{\m}_u\bar{\lambda}^{(\mathrm g)\ u}_{\ad v}\w^{\ad v}
  \notag\\
 &\quad +i\bar{\w}_{\ad \, u}\bar{\lambda}^{(\mathrm g)\ \ad u}_{v}
  \mu^v+\bar{\w}_{\ad \,\, u}\bar{\sigma}^{\m\n \ad\bd}(F_{\m\n})^u_v\w^{\bd \, v}
\end{align}
In this paper, we mainly investigate 
the contribution from the fermionic components 
of quarks, $\psi$ and $\tp$. 
We do not consider the terms with the gauge fields and gauginos, 
and $S_3$ indicates only terms with $\bar\psi$ or $\bar\tp$, hereafter. 

The space-time approach to obtain the superpotential or $F$-term
contribution 
from D-instanton was explained in \cite{Witten}. 
Following that, we should 
evaluate the following path integral for one-instanton 
\begin{equation}
 Z = C \int 
 d\{a,\chi,M,\lambda,D,\omega,\bar\omega,\mu,\bar\mu\}
  \ e^{-S_1-S_2-S_3} , 
\end{equation}
where, $C$ represents some overall constant. 
In the path integral evaluation,  
$a^{\m}$ gives rise to four bosonic translation modes $x^{\m}$ 
which represent position of the instanton. 
The upper left component of $M^{\a 4}$ 
gives rise to super-translation modes $\theta^{\a}$, 
since they are super-transform of $a_\mu$. 
Then, we can interpret the path integral over these fields 
as the integration with respect to (super-)coordinate. 
Then, the $F$-term contributions from D-instantons can be expressed as 
\begin{equation}
Z=\int d^4 x\, d^2\theta\ W , 
\end{equation}
where $W$ can be interpreted as a correction to the superpotential, 
and is given by 
\begin{equation}
 W=C\int d\{\lambda, D, \w,\bar{\w}, \m, \bar{\m}\}\ e^{-S_1-S_2-S_3}. 
  \label{Superpotential0}
\end{equation}
The integrals over $D$ and $\lambda$ enforces the bosonic and fermionic
ADHM constraints. If we carry out these integration we have
\begin{equation}
W=C \int d\{\w, \bar{\w}, \m, \bar{\m}\}
\delta(\bar{\m}_v\w^u_{\ad}+\bar{\w}_{\ad \, u}\m^v)
\delta(\bar{\w}^{\ad}_u(\tau^c)^{\bd}_{\,\,\ad}\w^u_{\bd})e^{-S_2-S_3}.
\end{equation}

In the computation of the ADS superpotential, one can consider only the
contribution
from $S_2$. If we consider the $SU(N_c)$ gauge theory, one can easily
see that $N_f=N_c-1$ is the condition for nonvanishing fermionic
integration from the structure of $S_2$. 
From $S_2$ one pair of $\m^u, \m^{\prime f}$ comes down to the
integrand simultaneously and
$\delta(\bar{\m}_v\w^u_{\ad}+\bar{\w}_{\ad \,u}\m^v)$
will give additional $\m^u$ to the integration. 
Thus we need $N_f=N_c-1$ for
the nonzero fermionic integration.
Now we consider the $N_f=N_c$ case. 
For nonzero fermionic integration, we
need $S_3$ to pull down two more fermion terms after pulling
down fermion terms up to $N_c-1$ from $S_2$. 
After carrying out these integration, 
we obtain the following expression: 
\begin{align}
 W = C\int d\omega^2d\bar\omega^2\ 
 \delta^{(3)}(\bar\omega_{\ad u}(\tau^c)^\ad_\bd\omega^{\bd u})\,
 \,&e^{-\frac{1}{2}\bar\omega_{\ad u} 
  \left(Q^u_f Q^{\dag f}_v 
   + \widetilde Q^{\dag u}_f \widetilde Q^f_v\right)\omega^{\ad v}}\notag\\
 \times\Biggl\{
 \frac{1}{2N!(N-2)!}
 &\biggl[\varepsilon_{uvt_1\cdots t_{N-2}}\varepsilon^{fgh_1\cdots h_{N-2}}
 \varepsilon^{s_1\cdots s_N}\varepsilon_{k_1\cdots k_N} 
 \notag\\&\qquad\times
 \omega_\ad^u\omega^{\ad v} 
 \bar\omega_{\bd x}\bar{\widetilde\psi}^{\bd x}_f 
 \bar\omega_{\gd y}\bar{\widetilde\psi}^{\gd y}_g 
 \widetilde Q^{\dag t_1}_{h_1} \cdots \widetilde Q^{\dag t_{N-2}}_{h_{N-2}}
 Q^{\dag k_1}_{s_1} \cdots Q^{\dag k_{N}}_{s_{N}}
 \bigg] \notag\\ 
 -\frac{2}{(N-1)!(N-1)!}
 &\biggl[\varepsilon_{ut_1\cdots t_{N-1}}\varepsilon^{fh_1\cdots h_{N-1}}
 \varepsilon^{vs_1\cdots s_{N-1}}\varepsilon_{gk_1\cdots k_{N-1}} 
 \notag\\&\qquad\times
 \omega_\ad^u\bar\omega^{\ad}_v 
 \bar\omega_{\bd x}\bar{\widetilde\psi}^{\bd x}_f 
 \bar{\psi}_{\gd y}^g \omega^{\gd y}
 \widetilde Q^{\dag t_1}_{h_1} \cdots \widetilde Q^{\dag t_{N-1}}_{h_{N-1}}
 Q^{\dag k_1}_{s_1} \cdots Q^{\dag k_{N-1}}_{s_{N-1}}
 \bigg] \notag\\ 
 +\frac{1}{2N!(N-2)!}
 &\biggl[\varepsilon_{t_1\cdots t_{N}}\varepsilon^{h_1\cdots h_{N}}
 \varepsilon^{uvs_1\cdots s_{N-2}}\varepsilon_{fgk_1\cdots k_{N-2}} 
 \notag\\&\qquad\times
 \omega_\ad^u\omega^{\ad v} 
 \bar\omega_{\bd x}\bar{\widetilde\psi}^{\bd x}_f 
 \bar\omega_{\gd y}\bar{\widetilde\psi}^{\gd y}_g 
 \widetilde Q^{\dag t_1}_{h_1} \cdots \widetilde Q^{\dag t_{N}}_{h_{N}}
 Q^{\dag k_1}_{s_1} \cdots Q^{\dag k_{N-2}}_{s_{N-2}}
 \bigg] 
 \Biggr\} ,  
\end{align}
where $N=N_f=N_c$. 
As in the case of the ADS superpotential, 
this expression represents 
lowest components of the effective superpotential. 
The fermions $\bar\psi_\ad$ and $\bar\tp_\ad$ 
can be regarded as the lowest component of 
the superfield $\bar D_\ad \bar Q$ 
and $\bar D_\ad \bar \tQ$, 
where $Q$'s are promoted to the superfields. 
Schematically, this is in the form of 
\begin{equation}
 \bar{\psi}_\ad\bar\psi^\ad F(Q,Q^\dagger)
  \to 
 \bar D_{\ad}\bar Q \bar D^{\ad}\bar Q F(Q,\bar Q) . \label{DMDM}
\end{equation}
This term is not the superpotential in the usual sense, 
and is not manifestly supersymmetric. 
However, this ``superpotential'' is 
chiral in the on-shell supersymmetry algebra because 
it is related to a representative of 
a Dolbeault cohomology 
whose elements parametrize infinitesimal 
deformations of moduli space \cite{Beasley:2004ys}. 
Since this interaction term contains a four-fermion interaction, 
it is called multi-fermion $F$-term.%
\footnote{
This type of the effective potential 
was also considered in \cite{Blumenhagen:2007bn}. 
}

Similarly one can clearly see that the nonvanishing contribution 
can be obtained even for $N_f > N_c$. 
We just have to pull down the necessary terms
from $S_3$. 
It is a formidable task to perform integration in general cases, 
and we consider the simplest cases. 
We will consider the case of SU(2) with $N_f=N_c$. 
In this case, the superpotential can be connected 
to the moduli deformation. 
In the next section, we review the relation between 
multi-fermion $F$-terms and the moduli space deformation. 

\section{Review on multi-fermion $F$-terms}\label{sec:Rev}

In this section, 
we briefly review 
the multi-fermion $F$-terms. 
The simplest multi-fermion $F$-term 
is obtained from 
the superpotential which has two fermions 
in the lowest component. 
In this case, 
the superpotential can be regarded as 
a correction to the K\"ahler potential 
and connected to the deformation of 
the moduli space of vacua. 

Let us consider a model 
whose moduli space is classically 
described by the complex fields $\phi^I$ satisfying 
a holomorphic equation, 
\begin{equation}
 \cC(\phi) = 0 . 
\end{equation}
The effective action is written in terms 
of $\phi$ as 
\begin{equation}
 S = \int d^4 x\,d^4\theta\,K(\Phi,\bar\Phi)
\end{equation}
where $\Phi^I$ and $\bar\Phi^{\bar I}$ are 
chiral and anti-chiral superfields whose 
lowest components are $\phi^I$ and 
$\bar\phi^{\bar I}$, respectively. 
Then, the bosonic part of this action is 
\begin{equation}
 S = \int d^4x\,g_{I\bar J}\partial\phi^I\partial\bar\phi^{\bar J} , 
\end{equation}
where $g_{I\bar J}$ is metric on the moduli space, 
and obtained from the K\"ahler potential as 
\begin{equation}
 g_{I\bar J} = \deriv{^2 K(\phi,\bar\phi)}{\phi^I\partial\bar\phi^{\bar J}} . 
\end{equation}

Suppose that the governing equation on the moduli space 
is deformed to 
\begin{equation}
 \cC(\phi) = \epsilon. 
\end{equation}
This deformation of the complex structure can be represented as the 
change of the basis of holomorphic one forms
\begin{align}
 d\phi^I &\to d\phi^I - \omega^I_{\bar J} d\bar\phi^{\bar J} . 
\end{align}
Then, the metric on the moduli space changes as 
\begin{equation}
 g_{I\bar J} d\phi^Id\bar\phi^{\bar J}
  \to g_{I\bar J} \left(d\phi^I - \omega^I_{\bar J} d\bar\phi^{\bar
                   J}\right)
  d\bar\phi^{\bar J} , 
\end{equation}
and the sigma model action receives the correction term 
\begin{align}
 \delta S &= \int d^4x\,d^2\theta\,
  \omega_{\bar I\bar J}\bar D_\ad\bar\Phi^{\bar I}
 \bar D^\ad\bar\Phi^{\bar J} 
  = \int d^4x\,\omega_{\bar I\bar J}
 \partial\bar\phi^{\bar I}\partial\bar\phi^{\bar J} +\cdots\\ 
 \omega_{\bar I\bar J} &= g_{K\bar I}\omega^K_{\bar J}. 
\end{align}
Thus, the deformation of the moduli space is 
related to the multi-fermion $F$-terms. 
Note that away from the singularities on the moduli space, 
this deformation can be converted to 
the non-holomorphic change of the valuable: 
\begin{equation}
 \phi^I \to \widetilde\phi^I = \phi^I + \delta\phi^I(\phi,\bar\phi), 
\end{equation}
where $\widetilde\phi^I$ satisfies the classical constraint 
\begin{equation}
 \cC(\widetilde\phi) = 0 . 
\end{equation}

In the next section, we work out 
the $SU(2)$ case and see that 
the result agrees with the computation 
carried out in \cite{Beasley:2004ys}. 

\vspace{5mm}
\section{Computation of multi-fermion $F$-terms}\label{sec:SU(2)}

In the previous section, 
we have seen that multi-fermion $F$-terms 
in the form of the integral with respect to instanton moduli. 
However, it is difficult to evaluate the integral for general $SU(N)$. 
In this section, we investigate 
the simplest case of multi-fermion $F$-terms, 
i.e. $SU(N)$ SQCD for $N_f=N_c=2$. 
In this case, multi-fermion $F$-terms 
are calculated in \cite{Beasley:2004ys}. 
We show that our D-instanton derivation 
correctly reproduces their result, 
which is related to the moduli space deformation. 

We start with the expression of \eqref{Superpotential0}, 
\begin{equation}
 W=C\int d\{\lambda, D, \w,\bar{\w}, \m, \bar{\m}\}\ e^{-S_1-S_2-S_3}. 
\end{equation}
We first perform integrations with respect to fermions. 
Integrals over $\lambda_\ad$ enforce the 
fermionic ADHM constraints. 
Here, we do not consider the terms with 
gauginos (and gauge fields). 
By neglecting these terms, 
integrals over $\mu$'s 
determine terms which contribute 
to the superpotential. 
After these integrations, 
we obtain the following expression of 
the superpotential, up to an overall integration constant: 
\begin{align}
 W &= \int d^2\omega\,d^2\bar\omega\,d^3D\ 
 \widetilde W\, 
 e^{-\frac{1}{2}\bar\omega(QQ^\dagger+\tQ^\dagger\tQ)\omega
 +iD^c\bar\omega\tau^c\omega} , \label{Superpotential2}\\
 \widetilde{W}&=\frac{1}{4}\varepsilon_{u_1u_2}\varepsilon^{f_1f_2}\w_{\ad}^{u_1}
 \w^{\ad \, u_2}\bar{\w}_{\bd \, x}{\bar{\tp}}^{\bd \, x}_{\,\,f_1}
 \bar{\w}_{\dot{\gamma}y}{\bar{\tp}}^{\dot{\gamma} \, y}_{\,\, f_2}
 \varepsilon^{v_1v_2}\varepsilon_{g_1g_2} 
 Q^{\dagger\,g_1}_{v_1}Q^{\dagger\,g_2}_{v_2} \notag \\
 &\quad + 2\varepsilon_{u_1u_2}
 \varepsilon^{f_1f_2}
 \w_{\ad}^{\,\, u_1}\tQ^{\dagger \,u_2}_{\,\,\, f_2}
 \bar{\w}_{\bd\, x}
 {\bar{\tp}}^{\bd \, x}_{\,\,f_1}
 \varepsilon^{v_1v_2}\varepsilon_{g_1g_2}
 \bar{\w}^{\ad}_{\,\,v_1}\bar{\psi}^{\gd \, g_1}_{\,\,\,y}
 \w_{\gd}^{\,\,y}Q^{\dagger \, g_2}_{\,\, v_2} \notag\\
 &\quad + \frac{1}{4}
 \varepsilon_{v_1v_2}\varepsilon^{g_1g_2}
 \tQ^{\dagger\,v_1}_{g_1}\tQ^{\dagger\,v_2}_{g_2}
 \varepsilon^{u_1u_2}\varepsilon_{f_1f_2}\bar{\w}_{\ad \, u_1}
 \bar{\w}^{\ad}_{\,\, u_2}\bar{\psi}^{\bd \, f_1}_{\,\,\, x}
 \w_{\bd}^{\,\, x} \bar{\psi}^{\gd \, f_2}_{\,\,\,y}\w^y_{\gd} . 
\end{align}
Using the Fierz identity%
\footnote{
For bosonic spinors
$A$, $B$, $C$ and $D$, 
we have the following relation: 
\begin{equation}
A_{\ad}B^{\ad}C_{\bd}D^{\bd}=(AB)(CD)=-(AD)(BC)+(AC)(BD) . 
\end{equation}
This can be used for color and flavor $SU(2)$ indices as well. 
}
for the first and the third terms, we obtain
an expression with one pair of $\w, \bar{\w}$ with their spinor indices
contracted, i.e., 
\begin{equation}
 (\w^u\w^v)(\bar{\w}_x\bar{\widetilde{\psi}}^x_f) 
  =-(\w^u\bar{\widetilde{\psi}}^x_f)(\w^v\bar{\w}_x)
  +  (\w^v\bar{\widetilde{\psi}}^x_f)(\w^u\bar{\w}_x) , \label{FierzSpinor}
\end{equation}
where spinor indices are contracted in each parenthesis. 
Then, $\widetilde W$ can be expressed as 
\begin{align}
 \widetilde{W}=\bar\omega_{\ad u_1}\omega^{\ad v_1}
 \bar\omega_{\bd u_2}\omega_\gd^{v_2}
 \Bigl[&\tfrac{1}{2}\varepsilon_{v_1v_2}\varepsilon^{f_1f_2}
 {\bar{\tp}}^{\bd u_1}_{\,\,f_1}
 {\bar{\tp}}^{\dot{\gamma} \, u_2}_{\,\, f_2}
 \varepsilon^{w_1w_2}\varepsilon_{g_1g_2} 
 Q^{\dagger\,g_1}_{w_1}Q^{\dagger\,g_2}_{w_2} \notag \\
 & - 2\varepsilon_{v_1w_1}
 \varepsilon^{f_1f_2}
 \tQ^{\dagger \,w_1}_{\,\,\, f_2}
 {\bar{\tp}}^{\bd \, u_2}_{\,\,f_1}
 \varepsilon^{u_1w_2}\varepsilon_{g_1g_2}
 \bar{\psi}^{\gd \, g_1}_{\,\,\,v_2}
 Q^{\dagger \, g_2}_{\,\, w_2} \notag\\
 &+ \tfrac{1}{2}
 \varepsilon_{w_1w_2}\varepsilon^{g_1g_2}
 \tQ^{\dagger\,w_1}_{g_1}\tQ^{\dagger\,w_2}_{g_2}
 \varepsilon^{u_1u_2}\varepsilon_{f_1f_2}
 \bar{\psi}^{\bd \, f_1}_{\,\,\, v_1}
 \bar{\psi}^{\gd \, f_2}_{\,\,\,v_2}\Bigr]. \label{Superpotential3}
\end{align}
Next, we perform the integration with respect to 
$\omega$ and $\bar\omega$. 
After some algebras, 
we obtain the following relation (see Appendix \ref{CGauss}): 
\begin{align}
 F(A) &= \int d^2\omega\,d^2\bar\omega\,d^3D\ 
 e^{-\bar\omega A\omega +iD^c\bar\omega\tau^c\omega} \notag\\
 &= \int d^3D \frac{1}{\det(A^2+D^2)} \notag\\
 &= (\tr A)^{-1} . \label{CGaussSU(2)}
\end{align}
where the matrix $A$ is a $2\times 2$ hermitian matrix, 
and given by 
$A=\frac{1}{2}(QQ^{\dagger} +\tQ^{\dagger}\tQ)$ in the present case. 
Picking up relevant parts in 
\eqref{Superpotential2} and \eqref{Superpotential3}, 
we obtain 
\begin{align}
 \int d\w\, d\bar{\w}\,d^3D\ \bar{\w}_{\ad u} \w^{\ad v} 
 \bar{\w}_{\bd \, u'}\w_{\gd}^{v'}
 e^{-\bar\omega A\omega +iD^c\bar\omega\tau^c\omega} 
 &= \frac{1}{2}\varepsilon_{\bd \gd} 
  \int d\w\, d\bar{\w}\,d^3D\ \bar{\w}_{\ad u} \w^{\ad v} 
 \bar{\w}_{\ad' u'}\w^{\ad'v'}
 e^{-\bar\omega A\omega +iD^c\bar\omega\tau^c\omega} \notag\\
 &=  \frac{1}{2}\varepsilon_{\bd \gd}\frac{\partial}{\partial A^u_v}
 \frac{\partial}{\partial A^{u'}_{v'}} F(A)
 =\varepsilon_{\bd \gd}\frac{\delta^v_u
 \delta^{v'}_{u'}}{(\tr A)^3} .
\end{align}
Using this, the superpotential is given by
\begin{align}
 W=(\tr A)^{-3}\Bigl(&
 -\frac{1}{2}\varepsilon^{uv}\varepsilon^{fg}\bar{\tp}^u_{\ad\,\,f} 
 \bar{\tp}^{\ad\,\,v}_f
 \varepsilon^{u'v'}\varepsilon_{f'g'}Q^{f'}_{u'}Q^{g'}_{v'} \notag\\
 &+2\varepsilon_{ut}\varepsilon^{fk}\varepsilon^{us}\varepsilon_{gk}   
 \bar{\tp}^x_{\ad\,\, f}\bar{\psi}^{\ad\,\,g}_x  
 \tQ^{\dagger \, t}_{\,\, k}Q^{\dagger \, k}_{\,\, s} \notag\\
 &-\frac{1}{2}\varepsilon^{uv}\varepsilon^{fg}
 \tQ^{\dagger\ u}_{f}\tQ^{\dagger\ v}_f
 \varepsilon^{u'v'}\varepsilon_{f'g'}
 \bar\psi^{f'}_{\ad u'}\bar\psi^{\ad g'}_{v'} \Bigr) . 
\end{align}

In the case of $SU(2)$, 
the fundamental and anti-fundamental representations coincide, 
and the $N_f$ flavors can be treated as $2N_f$ flavors. 
Here, we treat the flavor symmetry as global symmetry 
by neglecting gauge fields of flavor symmetry. 
Then, the flavor symmetry becomes $SU(4)$. 
First, we rewrite the quarks $\tQ$ in the fundamental 
representation of $SU(2)$ color symmetry as 
\begin{align}
 Q^{\prime uf} &= \varepsilon^{uv}\tQ^{f}_v , &
 \psi_\alpha^{\prime uf} &= \varepsilon^{uv}\tp^{f}_{\alpha v} . 
\end{align}
By applying the the Fierz identity to color $SU(2)$ indices, 
each term in the parenthesis becomes 
\begin{align}
 &-\varepsilon^{fg}\varepsilon_{f'g'}(Q^{\prime\dagger \,f'}\bar{\psi}_f)  
 (Q^{\prime\dagger g'}\bar{\psi}_g), \\
 &2\varepsilon^{fg}\varepsilon_{f'g'}  
 ((\bar{\psi}_fQ^{\dagger}_g)(\bar{\psi}^{\prime f'}Q^{\prime \dagger g'})  
 + (\bar{\psi}_fQ^{\prime \dagger f'})
 (\bar{\psi}^{\prime g'}Q^{\dagger}_{g})), \\
 &-\varepsilon^{fg}\varepsilon_{f'g'} (Q^{\prime\dagger \,f'}\bar{\psi}_{f})  
 (Q^{\prime\dagger g'}\bar{\psi}_{g}) ,
\end{align}
respectively, 
where $SU(2)$ color indices are contracted inside 
each parenthesis and are not written explicitly. 
Thus, the superpotential is given by 
\begin{align}
 W = (\tr A)^{-3}\varepsilon^{fg}\varepsilon_{f'g'}
 \Bigl[&-(Q^{\prime\dagger \,f'}\bar{\psi}_f)  
 (Q^{\prime\dagger g'}\bar{\psi}_g) \notag\\
 &+2 \left((\bar{\psi}_fQ^{\dagger}_g)
 (\bar{\psi}^{\prime f'}Q^{\prime \dagger g'})  
 + (\bar{\psi}_fQ^{\prime \dagger f'})
 (\bar{\psi}^{\prime g'}Q^{\dagger}_{g})\right) \notag\\
 &-(Q^{\prime\dagger \,f'}\bar{\psi}_{f})  
 (Q^{\prime\dagger g'}\bar{\psi}_{g}) \Bigr] . 
\end{align}
By promoting the quarks $Q$ and their superpartners $\psi$ to the superfields, 
fermion $\bar\psi_\ad$ can be interpreted as 
the lowest component of $\bar D_\ad \bar Q$. 
The quarks can be combined into $\cQ$ as 
\begin{equation}
 \cQ^u_i = 
  \begin{cases}
   Q^u_f , & (i=1,2) \\
   Q^{\prime uf}  . & (i=3,4)
  \end{cases}
\end{equation}
Then, the superpotential can be expressed in the following 
compact form 
\begin{equation}
 W = (\tr A)^{-3} \varepsilon_{ijkl}
  \bar D_\ad (\varepsilon^{uv}\bar\cQ_u^i\bar\cQ_v^j) 
  \bar D^\ad (\varepsilon^{u'v'}\bar\cQ_{u'}^k\bar\cQ_{v'}^l) , 
\end{equation}
where the matrix $A$ can be written as $A = \cQ\bar\cQ$. 
In the moduli space of SQCD, 
we have the $D$-flatness condition, $D^a =\tr\cQ^{\dagger}\tau^a\cQ=0$,  
and we can rewrite $\tr A$ as 
\begin{equation}
 \tr A = \sqrt{\det\bar\cQ\cQ} . 
\end{equation}
Introducing the ``meson'' field $M_{ij} = \varepsilon_{uv}\cQ^u_i\cQ^v_j$, 
we obtain the following expression for the superpotential 
\begin{equation}
 W=(\tr\bar{M}M)^{-\frac{3}{2}}\varepsilon^{ijkl}
  \bar{D}_\ad\bar{M}_{ij}\bar{D}^\ad\bar{M}_{kl} .  \label{SuperpotentialMeson}
\end{equation}

In \cite{Beasley:2004ys}, 
the superpotential is derived by the usual field theory instanton calculation
and its relation to the moduli space deformation is explained. 
In the case of $SU(2)$, 
the classical moduli space is described 
by the meson $M$ satisfying the classical constraint 
\begin{equation}
 \Pf M = 0 . 
\end{equation}
This constraint is modified 
by the quantum effect to 
\begin{equation}
 \Pf M = \epsilon . 
\end{equation}
This quantum moduli space can be converted 
to the classical moduli space by a non-holomorphic 
change of variables. 
Introducing the new coordinate $\tM = M - \delta M$, 
the deformation of the complex structure 
gives rise to the additional superpotential 
of the form 
\begin{equation}\label{MesonSuperpotential}
 \omega_{ijkl} G^{ij}_{mn}\bar D\bar M^{kl} \bar D \bar M^{mn} , 
\end{equation}
where $G^{ij}_{kl}$ is the metric of the moduli space, 
and $\omega_{ijkl}$ is given by 
\begin{equation}
 \left(\deriv{}{\bar M^{kl}} + \omega_{ijkl}\deriv{}{M_{ij}}\right)
  \tM_{mn} = 0 . 
\end{equation}
In this case, the new coordinate is given by 
\begin{equation}
 M_{ij} \to M_{ij} 
  -\frac{\epsilon}{2}\frac{\varepsilon_{ijkl}\bar M^{kl}}{(\tr \bar M M)} . 
\end{equation}
Then, $\omega_{ijkl}$ becomes 
\begin{equation}
 \omega_{ijkl} = \frac{\epsilon}{2}\left(
  \frac{\varepsilon_{ijkl}}{(\tr\bar{M}M)}
  -\frac{\varepsilon_{ijmn}\bar{M}^{mn}M_{kl}}{(\tr\bar{M}M)^2}\right) . 
\end{equation}
The metric can be determined 
by the asymptotic form of the K\"ahler potential 
$K = \sqrt{\tr \bar M M}$. 
The additional factor of $(\tr \bar M M)^{-1/2}$ 
comes from this metric.%
\footnote{
There are another additional terms. 
We do not write these terms explicitly 
because they vanish on the moduli space.  
}
Then, the superpotential becomes, 
\begin{equation}
 W=(\tr\bar{M}M)^{-\frac{3}{2}}
  \left(\varepsilon_{ijkl}
   -\frac{\varepsilon_{ijmn}\bar{M}^{mn}M_{kl}}{(\tr\bar{M}M)} 
  -\frac{\varepsilon_{mnkl}\bar{M}^{mn}M_{ij}}{(\tr\bar{M}M)}\right)
  \bar{D}\bar{M}^{ij}\bar{D}\bar{M}^{kl} , 
\label{bw}
\end{equation}
where the third term has been added so that 
the superpotential is manifestly symmetric 
to the exchange of two $\bar D \bar M$'s. 
It can be done because 
the second and third term vanish on the moduli space since 
\begin{equation}
 \varepsilon^{ijkl}\bar{M}_{kl}d\bar{M}_{ij}
  =d(\varepsilon^{ijkl}\bar{M}_{ij}\bar M_{kl}) . \label{VanishOnModuli}
\end{equation}
Using this relation \eqref{VanishOnModuli} again, 
the superpotential \eqref{SuperpotentialMeson} 
is equivalent to \eqref{bw} which is based on the 
moduli space deformation. 

\section{Case of symplectic gauge group}\label{sec:USp}

In this section, we consider the case of 
the symplectic gauge group, 
which can be obtained by introducing the orientifold 
\cite{Douglas:1996sw}. 
Here, we take the orbifolding with 
$(2N_c,2N_f,0,0)$ for D3-branes and 
$(1,0,0,0)$ for D-instantons, 
and put the O3-plane on the orbifold singularity. 
Then, the color and flavor gauge groups becomes 
$USp(2N_c)$ and $USp(2N_f)$, respectively. 
If one neglects the flavor gauge symmetry 
and treats this as a global symmetry, 
this remains $U(2N_f)$. 

The orientifold acts on the Chan-Paton factor 
as a matrix $\gamma(\Omega)$. 
This matrix can be different for different D-branes, 
and have opposite symmetries (symmetric or anti-symmetric) 
for the D3-brane and D-instanton \cite{Gimon:1996rq}. 
We take the anti-symmetric one for the D3-brane 
to obtain the symplectic gauge group. 
The matrix $\gamma(\Omega)$ must satisfy 
the following consistency condition: 
\begin{equation}
 \gamma(g)\gamma(\Omega)\gamma(g) = \gamma(\Omega) . 
\end{equation}
Then, the action of the orientifold is 
\begin{equation}
 \gamma_- = \left(
  \begin{matrix}
   J & 0 & 0 & 0 \\
   0 & \widetilde J & 0 & 0 \\
   0 & 0 & J^{(3)} & 0 \\
   0 & 0 & 0 & J^{(4)} 
  \end{matrix}\right) , 
\end{equation}
where, $J$'s are anti-symmetric matrices 
which satisfies $J^2 = -1$. 
We use the notation of 
$J_{uv} = - J^{uv} = (J^{-1})_{uv}$, and similarly for $\tJ$. 

The orientifold imposes the following additional 
conditions on massless fields in the D3-D3 sector: 
\begin{subequations}\label{O3D3}
 \begin{align}
  A_\mu &= - \gamma_- A_\mu^\rT \gamma_-^{-1} , \\
  \Phi &= - \gamma_- \Phi^\rT \gamma_-^{-1} , & 
  \Phi^\dagger &= - \gamma_- \Phi^{\dagger\,\rT} \gamma_-^{-1} , \\
  \Psi^A &= - R^A_B \gamma_- (\Psi^B)^\rT \gamma_-^{-1} , & 
  \bar\Psi_A &= \gamma_- (\bar\Psi_B)^\rT \gamma_-^{-1} R^B_A ,  
 \end{align}
\end{subequations}
where the minus sign for the scalars is due to the 
spacetime reflection, and 
those for the gauge fields and the fermions come from 
the worldsheet reflection. 
The matrix $R^A_B$ is the action of the reflection 
on six dimensional spinors: 
\begin{equation}
 R = -i \Gamma^{456789} . 
\end{equation}
The projection on the gauge fields (and gauginos) 
restrict the color and flavor gauge groups 
to the symplectic gauge group. 
For the chiral scalars and fermions, 
the conditions \eqref{O3D3} make connections between 
$Q$ and $\widetilde Q$, $\psi$ and $\widetilde\psi$ as 
\begin{align}
 \tQ^f_u &= - \tJ^{fg} Q^v_g J_{vu} , & 
 \tQ^{\dag u}_f &= - J^{uv} Q^{\dag g}_v \tJ_{gf} , & 
 \tp^f_u &= - \tJ^{fg} \psi^v_g J_{vu} , &
 \bar\tp^u_f &= - J^{uv} \bar\psi_v^g \tJ_{gf} . 
\end{align}
Since these conditions do not give any 
restriction on the flavor symmetry, 
the global symmetry itself remains to be $U(2N_f)$. 

For the Chan-Paton factor on the D-instanton, 
the action of the orientifold is 
a symmetric matrix, which can be taken to be 
the unit matrix 
\begin{equation}
 \gamma_+ = 1 . 
\end{equation}
For the fields in the D(-1)-D(-1) sector, 
the orientifold imposes the following conditions: 
\begin{align}
 a_\mu &= a_\mu^\rT , & 
 \chi &= - \chi^\rT , & 
 M_\alpha^A &= R^A_B (M_\alpha^B)^\rT , & 
 \lambda_{\ad A} &= (\lambda_{\ad B})^\rT R^B_A . 
\end{align}
After the orbifolding for 
the one-instanton condition, (1,0,0,0), 
these relations yield 
\begin{align}
 \lambda_\ad &= - \lambda_\ad , & 
 D^c &= - D^c. 
\end{align}
Thus, $\lambda$ and $D^c$ must vanish. 
It implies that there are no bosonic and fermionic 
ADHM constraints in $USp(2N_c)$ gauge theory as is well-known. 

For the fields in the D3-D(-1) sector, 
the orientifold projection gives relations between 
$\omega$ and $\bar\omega$, and, $\mu$ and $\bar\mu$: 
\begin{align}
 \bar\omega &= \omega^\rT \gamma_-^{-1} , &
 \bar\mu^A &= R^A_B (\mu^B)^\rT \gamma_-^{-1}. 
\end{align}
These relations become the following relations 
after the orbifolding: 
\begin{align}
 \bar\omega_u &= \omega^v J_{vu} , &
 \bar\mu_u &= \mu^v J_{vu} , & 
 \bar\mu'_f &= \mu^g \tJ_{gf} .
\end{align}

The superpotential is 
greatly simplified since 
the ADHM constraints are absent in this case. 
The action of our interests 
becomes 
\begin{equation}\label{UspAction}
 S = \omega_{\ad}^u J_{uv} Q^v_f Q^{\dagger f}_w \omega^{\ad w} 
  + i \mu^{\prime f} \tJ_{fg} Q^{\dagger g}_u \mu^u 
  + 2 i \mu^{\prime f} \tJ_{fg} \bar\psi^g_{\ad u} \omega^{\ad u} . 
\end{equation}

\subsection{Multi-fermion $F$-terms and the moduli space deformation}

In this section, we consider the simplest case 
of the multi-fermion $F$-term. 
As we have seen in the case of $SU(2)$, 
the superpotential is related to 
the deformation of the moduli space. 
We generalize the analysis to the symplectic case, 
and show that our superpotential reproduces 
the moduli deformation. 

The superpotential $W$ is given by the path integral of 
the action \eqref{UspAction}: 
\begin{equation}
 W = \int d\{\omega,\mu\} e^{-S} . 
\end{equation}
We neglect contributions from gauginos. 
Then, all of $\mu$ in the integrand must 
come from the second term of \eqref{UspAction}, 
because the ADHM constraints are absent in this case. 
If $N_c = N_f$, there are no additional contribution of $\mu'$, 
and it yields an ADS-type superpotential, as is well known 
\cite{Intriligator:1995ne}. 
In the case of $N_f > N_c$, $N_c$ of $\mu'$ are 
supplied by the second term of \eqref{UspAction}, 
and the rest of $\mu'$ are by the third term. 
Then, the superpotential has the $2(N_f-N_c)$ of the fermions $\bar\psi$. 
We obtain the simplest multi-fermion $F$-terms 
in the case of $N_f = N_c + 1$. 
In this case, we have the follwing expression up to an overall constant: 
\begin{align}
 W &=  \int d^2\omega\ e^{-\omega_\ad (JQQ^\dagger) \omega^\ad}\notag\\ 
 &\qquad\qquad\times
 \varepsilon^{u_1\cdots u_{N}}\varepsilon^{f_1\cdots f_{N+2}} 
 (JQ^{\dagger})_{u_1 f_1}\cdots (JQ^{\dagger})_{u_{N}f_{N}}
 (\tJ\bar\psi\omega)_{f_{N+1}} (\tJ\bar\psi\omega)_{f_{N+2}} 
 \notag\\
 &=  \frac{1}{\det(JQQ^\dagger)}
 \varepsilon^{u_1\cdots u_{2N_c}}\varepsilon^{f_1\cdots f_{2N_f}}
 (JQ^{\dagger})_{u_1 f_1}\cdots (JQ^{\dagger})_{u_{2N_c}f_{2N_c}} \notag\\
 & \qquad\qquad \times
 (\tJ\bar\psi)_{\ad v f_{N+1}}\cdots 
 (\tJ\bar\psi)^{\ad}_{w f_{N+2}}
 \left((QQ^\dagger)J\right)^{vw}
\end{align}
where, $N = 2N_c = 2(N_f-1)$. 
Using the Bose statistics of the quarks $Q$, 
we can rewrite this superpotential 
in terms of the ``mesons'' $M_{fg} = J_{uv}Q^u_fQ^v_g$. 
Then, we obtain the following compact form of the superpotential: 
\begin{align}
 W &=  \frac{\bar C'(\bar M)_{ijkl}}{\bar C(\bar M)_{fg}C(M)^{fg}} 
  \widetilde G^{ij}_{mn}\bar D\bar M^{kl} \bar D \bar M^{mn} , 
 \label{UspSuperpotential} \\
 \widetilde G^{ij}_{kl} 
 &= Q^{\dagger i}_u \left((QQ^\dagger)^{-2}\right)^u_v Q^{v}_k\,\delta^j_l ,
\end{align}
where we define $C$, $\bar C$ and $\bar C'$ as 
\begin{align}
 C(M)^{ij} &= 
 \varepsilon^{ijf_1\cdots f_N} M_{f_1f_2}\cdots M_{f_{N-1}f_N} ,\label{DefC} \\
 \bar C(\bar M)_{ij} &= 
 \varepsilon_{ijf_1\cdots f_N} \bar M^{f_1f_2}\cdots \bar M^{f_{N-1}f_N} , \label{DefbarC}
\end{align}
and 
\begin{equation}
 \bar C'(\bar M)_{ijkl} = 
 \varepsilon_{ijklf_1\cdots f_{N-2}} 
 \bar M^{f_1f_2}\cdots \bar M^{f_{N-3}f_{N-2}} . \label{DefC'}
\end{equation}

Let us consider the relation between our superpotential 
and the moduli space deformation \cite{Intriligator:1995ne}. 
We treat only the color symmetry as the gauge symmetry, 
but the flavor symmetry as a global symmetry 
by neglecting the flavor gauge field. 
Then, our model has $USp(2N_c)$ color gauge symmetry and 
$U(2N_f)$ flavor global symmetry. 
The classical moduli space is described by the ``mesons'' $M_{fg}$. 
For $N_f > N_c$, these mesons satisfy the classical 
constraint 
\begin{equation}
 \Pf M = 0 , 
\end{equation}
which is a trivial consequence of the Bose statistics 
of the quark fields $Q$. 
In the case of $N_f = N_c + 1$, 
this constraint is modified to 
\begin{equation}
 \Pf M = \Lambda
\end{equation}
by the quantum effect of instantons. 

The quantum moduli space can be converted 
to the classical moduli space by a non-holomorphic 
change of variables. 
Introducing the new coordinates $\tM = M - \delta M$, 
The deformation of the complex structure 
gives rise to the additional superpotential 
of the form 
\begin{equation}\label{UspMesonSuperpotential}
 \omega_{ijkl} G^{ij}_{mn}\bar D\bar M^{kl} \bar D \bar M^{mn} , 
\end{equation}
where $G^{ij}_{kl}$ is the metric of the moduli space, 
and $\omega_{ijkl}$ is given by 
\begin{equation}
 \left(\deriv{}{\bar M^{kl}} + \omega_{ijkl}\deriv{}{M_{ij}}\right)
  \tM_{mn} = 0 . 
\end{equation}
In this case, the moduli deformation is described by 
\begin{equation}
 \delta M_{ij} = \frac{\bar C(\bar M)_{ij}}{\bar C(\bar M)_{fg}C(M)^{fg}} , 
\end{equation}
where we have used the definitions of 
\eqref{DefC} and \eqref{DefbarC} again. 
Then, we have 
\begin{equation}\label{UspOmega}
 \omega_{ijkl} = \frac{\bar C'(\bar M)_{ijkl}}{\bar C(\bar M)_{fg}C(M)^{fg}} 
  -\frac{\bar C(\bar M)_{ij}\bar C'(\bar M)_{klmn}C(M)^{mn}}
  {\left(\bar C(\bar M)_{fg}C(M)^{fg}\right)^2} ,  
\end{equation}
where $C'$ is defined in \eqref{DefC'}. 
We do not know the explicit form of the metric $G$ but 
we can estimate it from the asymptotic form of 
the K\"ahler potential $K$. 
The K\"ahler potential for mesons $M$
equals to that of quarks $Q$ at the asymptotic region of 
the moduli space. 
Since the $D$-flatness condition for 
the symplectic gauge group is 
\begin{equation}
 J_{uw}Q^w_fQ^{\dagger f}_v = Q^{\dagger f}_uQ^w_f J_{wv} , 
\end{equation}
the K\"ahler potential can be written as 
\begin{equation}
 K = \tr Q^\dagger Q 
  = \tr \sqrt{JQ^\rT Q^{\dagger\,\rT}J^{-1}QQ^{\dagger}} . 
\end{equation}
We can read off an expression of the metric 
from the last expression by 
using the following relation: 
\begin{align}
 \deriv{^2K}{Q^{\dagger g}_v\partial Q^u_f} 
  &= \deriv{^2K}{\bar M^{kl}\partial M_{ij}}
  \deriv{M_{ij}}{Q^u_f}\deriv{\bar M^{kl}}{Q^{\dagger g}_v} \notag\\
  &= G^{ij}_{kl}
  \deriv{M_{ij}}{Q^u_f}\deriv{\bar M^{kl}}{Q^{\dagger g}_v} . 
\end{align}
Thus, we obtain 
\begin{align}
 G^{ij}_{kl} &= 
 \delta^j_k\bar B^i_l + \bar B^j_k \delta^i_l 
 -\dfrac{1}{2}(D^j_k\bar B^i_l + \bar B^j_k D^i_l ) \notag\\ \label{UspMetric} 
 &\qquad +\dfrac12\sum_{r=1}^\infty (-1)^r 
 \left[(B^r)^j_k(\bar B^{r+1})^i_l + (\bar B^{r+1})^j_k (B^r)^i_l \right]
\end{align}
where 
\begin{align}
 B^i_j &= Q^{\dagger i}_uQ^{u}_j , & 
 \bar B ^i_j
 &= Q^{\dagger i}_u \left((QQ^\dagger)^{-2}\right)^u_v Q^{v}_j , & 
 D^i_j
 &= Q^{\dagger i}_u \left((QQ^\dagger)^{-1}\right)^u_v Q^{v}_j . & 
\end{align}
Substituting \eqref{UspOmega} and \eqref{UspMetric} into 
\eqref{UspMesonSuperpotential}, 
we obtain the superpotential generated by 
the quantum effect of instantons. 
Using the relation $M_{fg} = J_{uv}Q^u_fQ^v_g$ 
and the Bose statistics of $Q$, 
this superpotential is simplified to the following form: 
\begin{align}
 W &=  \frac{\bar C'(\bar M)_{ijkl}}{\bar C(\bar M)_{fg}C(M)^{fg}} 
  \widetilde G^{ij}_{mn}\bar D\bar M^{kl} \bar D \bar M^{mn} , \\
 \widetilde G^{ij}_{kl} &= \bar B^i_k \delta^j_l ,
\end{align}
which is equivalent to our superpotential \eqref{UspSuperpotential}. 

\subsection{Higher multi-fermion $F$-terms}

In the case of $N_f > N_c +1$, 
the moduli space is not deformed by effects of instantons, 
and we cannot relate multi-fermion $F$-terms to the moduli space deformation. 
However, multi-fermion $F$-terms themselves exist even in these cases. 
These multi-fermion $F$-terms are 
also found from the purely field theoretical analysis. 
Here, we restrict ourselves to the simplest 
gauge group of $USp(2) \simeq SU(2)$, 
and show that our analysis reproduces 
the result from field theory. 

In general cases of $N_f > N_c$, 
we have the following expression of the superpotential: 
\begin{align}\label{UspFterm}
 W &=  \int d^2\omega\ e^{-\omega_\ad (JQQ^\dagger) \omega^\ad}\notag\\ 
 &\qquad\qquad\times
 \varepsilon^{u_1\cdots u_{2N_c}}\varepsilon^{f_1\cdots f_{2N_f}} 
 (JQ^{\dagger})_{u_1 f_1}\cdots (JQ^{\dagger})_{u_{2N_c}f_{2N_c}}
 (\tJ\bar\psi\omega)_{f_{2N_c+1}}\cdots (\tJ\bar\psi\omega)_{f_{2N_f}} 
 \notag\\
 &=  \sum_\sigma \frac{1}{\det(JQQ^\dagger)}
 \varepsilon^{u_1\cdots u_{2N_c}}\varepsilon^{f_1\cdots f_{2N_f}}
 (JQ^{\dagger})_{u_1 f_1}\cdots (JQ^{\dagger})_{u_{2N_c}f_{2N_c}} \notag\\
 & \qquad\qquad \times
 (\tJ\bar\psi)_{\ad_1v_1f_{2N_c+1}}\cdots 
 (\tJ\bar\psi)_{\ad_{N_f-N_c}v_{N_f-N_c}f_{N_f+N_c}}
 (\tJ\bar\psi)^{\ad_1}_{w_1f_{N_f+N_c+1}}\cdots 
 (\tJ\bar\psi\omega)^{\ad_{N_f-N_c}}_{w_{N_f-N_c}f_{2N_f}} \notag\\
 & \qquad\qquad \times
 \left((QQ^\dagger)J\right)^{v_1\sigma(w_1)}\cdots
 \left((QQ^\dagger)J\right)^{v_{N_f-N_c}\sigma(w_{N_f-N_c})}. 
\end{align}
where, $\sigma$ indicates the permutation of indices, 
which can be removed by using the Fierz identity. 
Using the relation $M_{fg} = Q^u_f Q^v_g$ and the Bose 
statistics of quarks $Q$, 
we can simplify it into the following compact form: 
\begin{equation}
 W =  \frac{1}{\det(JQQ^\dagger)}\varepsilon_{f_1\cdots f_{2N_f}}
  \bar M^{f_1f_2}\cdots \bar M^{f_{2N_c-1}f_{2N_c}}
  \cO^{f_{2N_c+1}f_{2N_c+2}}\cdots \cO^{f_{2N_f-1}f_{2N_f}} , 
\end{equation}
where 
\begin{equation}
 \cO^{fg} = \bar\psi^f_{\ad u}\left[(QQ^\dagger)^{-1}\right]^u_w
  J^{wv}\bar\psi^{\ad g}_v . 
\end{equation}
Using the Bose statistics of $Q$ again, and extending $\bar\psi$ 
to the superfield $\bar D\bar Q$, 
we can replace $\cO$ by $\widetilde \cO$ which is defined as 
\begin{equation}
 \widetilde \cO^{fg} = \bar D \bar M^{ff'}
  \left[Q^\rT J (QQ^\dagger)^{-1}\right]_{f'g'}\bar D\bar M^{g'g} . 
\end{equation}
In the case of $USp(2) = SU(2)$, 
$D$-flatness condition gives the following relation: 
\begin{equation}
 (QQ^\dagger)^u_v = \delta^u_v \tr (QQ^\dagger) 
  = \delta^u_v \sqrt{\tr \bar M M} . 
\end{equation}
Then, we obtain 
\begin{equation}
 \widetilde \cO^{fg} = 
  (\tr \bar M M) ^{-3/2} M_{f'g'} \bar D \bar M^{ff'} \bar D \bar M^{gg'} , 
\end{equation}
Using this expression, we can rewrite the superpotential as 
\begin{align}
 W &=  \left(\tr\bar MM\right)^{-(3n-1)/2} \varepsilon_{f_1\cdots f_{2N_f}} 
 \bar M^{f_1f_2}\bar\cO^{f_3f_4}\cdots\bar\cO^{f_{2N_f-1}f_{2N_f}} , \\
 \bar\cO^{ij} &= M_{kl}\bar D_\ad \bar M^{ik} \bar D^\ad\bar M^{lj} . 
\end{align}
Therefore, this multi-fermion $F$-term is 
equivalent to that in \cite{Beasley:2004ys}.

\section{Conclusions and discussions}\label{sec:Concl}

In this paper, we have investigated 
the instanton in $\cN = 1$ SQCD 
by using the D-brane effective theory. 
In SQCD with gauge group $SU(N_c)$ and 
$N_f=N_c$ flavors, 
instantons modify the moduli space of vacua. 
This effect can be described 
by the multi-fermion $F$-terms, 
and we have derived these terms 
from the D-brane effective action. 
SQCD can be obtained by 
introducing the orbifolding 
into 
the D3-brane effective theory, 
and the gauge instanton corresponds to the 
D-instanton on the D3-brane. 
The effective potential generated by instantons can be 
obtained from the D-instanton effective action. 
The multi-fermion $F$-terms are given in terms of 
integral with respect to the instanton moduli. 
This integration generally gives 
complicated expression. 
We have considered the simplest case of $SU(2)$ gauge group, 
for which the potential was calculated 
in \cite{Beasley:2004ys} by using 
purely field theoretical techniques. 
Our result correctly reproduced that in \cite{Beasley:2004ys}. 
We have also considered the case of 
symplectic gauge group. 
In this case, we have obtained 
much simpler results 
than those in the case of the unitary group. 
This is due to the fact that 
the ADHM constraints are 
absent for the symplectic gauge group. 
We have shown that 
the deformation of the moduli space 
is described by the multi-fermion $F$-terms 
derived from D-instantons, for $N_f=N_c+1$. 
We have also calculated 
multi-fermion $F$-terms with more fermions, 
which appear for the theory with more flavors. 
For $USp(2)\sim SU(2)$, 
our result agrees with that in \cite{Beasley:2004ys}, again. 

We would like to comment on the case of the orthogonal group. 
SQCD with this gauge group can be obtained by 
introducing the orientifold which 
is opposite to that for the symplectic group; 
symmetric projection is imposed on the Chan-Paton factor 
of D3-branes and the anti-symmetric projection on that of D-instantons. 
We can obtain multi-fermion $F$-terms 
similar to those in the case of unitary gauge group 
for $N_f=N_c$ or symplectic gauge group 
for $N_f=N_c+1$, i.e. those in the form of \eqref{DMDM}.
Due to the anti-symmetric projection, 
the size of the matrices for D-instantons must be even, 
and we should take $k=2$ for one-instanton. 
Then, the Chan-Paton factor for D-instantons 
becomes $USp(2)\sim SU(2)$, 
and consequently, there are three sets of 
the ADHM constraints corresponding to three $SU(2)$ generators. 
The fermionic ADHM constraints supply 
six fermions $\mu$. 
Since there are $2N_c$ fermions $\mu$ 
and $2N_f$ fermions $\mu'$, 
we obtain the multi-fermion $F$-terms 
in the form of \eqref{DMDM}, if $N_f=N_c-2$. 
However, there are no constraint on the moduli space 
for $N_f=N_c-2$, and therefore, 
we cannot see any relation between 
the multi-fermion $F$-terms 
and the deformation of the moduli space \cite{Intriligator:1995id}. 
Furthermore, a large number of the ADHM constraint makes 
the integration in the expression of 
multi-fermion $F$-terms much more complicated. 

It would be interesting to study the relation between 
multi-fermion $F$-terms and the deformation of the moduli space 
in the case of $SU(N_c)$ for $N_c>2$. 
In order to study this relation, 
we have to obtain an explicit form of 
the metric on the moduli space. 
Even in the case of $SU(2)$ and symplectic group, 
the metric is determined 
in the asymptotic region by using symmetries. 
It is interesting to describe deformation of 
the moduli space in full detail. 
These issues are left for future studies. 

Another interesting problem is 
a generalization to other models. 
In this paper, 
we have demonstrated the D-instanton derivation of 
multi-fermion $F$-terms which are related 
to the deformation of the moduli space. 
This method can be applied to 
other models and will show how 
the moduli space is deformed by instantons. 
It would also be interesting to 
study stringy effects of multi-fermion $F$-terms.

\section*{Acknowledgments}

 J. Park thanks A. Uranga for the discussion where the current project
was initiated and for the collaboration in the initial stage of the project. 
We also would like to thank T.~Tsukioka for fruitful discussions. 
J. Park is supported by the Science Research Center Program
of KOSEF through the Center for Quantum Space-Time (CQUeST) of
Sogang-University with the grant number R11-2005-021, by the Postech
BSRI research fund 2007 and by the Stanford Institute for Theoretical Physics.

\appendix

\section{Supersymmetry transformations}\label{SUSY}

In this appendix, we describe the supersymmetry transformation 
of the effective theory. 
The D3-brane effective theory is 
the $\cN = 4$ super Yang-Mills theory. 
The followings are their transformations: 
\begin{align}
 \delta A_\mu &= i\bar\xi_{\ad A}\bar\sigma^{\mu\ad\beta}\Psi_{\beta}^A
 + \xi^{\alpha A} \sigma^\mu_{\alpha\bd} \bar\Psi_A^\bd , \\
 \delta \Psi^{\alpha A} 
 &= \tfrac{i}{2}\sigma^{\mu\nu\alpha}_\beta \xi^{\beta A} F_{\mu\nu} 
 +\tfrac{i}{2}\varepsilon^{ABCD}\bar\xi_{\bd B} 
 \bar\sigma^{\mu\bd\alpha} D_\mu\Phi_{AB} 
 + \left(\tfrac18\xi^{\alpha A}\varepsilon^{BCDE}
 -\tfrac12 \varepsilon^{ABCD}\xi^{\alpha E}\right) \Phi_{BC}\Phi_{DE} , \\
 \delta \bar\Psi_{\ad A} 
 &= \tfrac{i}{2}\bar\xi_{\bd A}\bar\sigma^{\mu\nu\bd}_\ad F_{\mu\nu} 
 +i\xi_{\beta B} \sigma^{\mu}_{\beta\ad} D_\mu\Phi_{AB} 
 + \varepsilon^{ABCD}\left(\tfrac18\bar\xi_{\ad A}\Phi_{BC}\Phi_{DE}
 -\tfrac12 \bar\xi_{\ad E}\Phi_{AB}\Phi_{CD}\right) , \\ 
 \delta \Phi_{AB} &= 2i\varepsilon_{ABCD} \xi^{C}_\alpha \Psi^{\alpha D} 
 -2i\left(\bar\xi_{\ad A} \bar\Psi^\ad_B -\bar\xi_{\ad B}\bar\Psi^\ad_A\right) . 
\end{align}
where, we rewrote the $R$-symmetry in terms of $SU(4)$ 
using the definition, $\Phi_{AB} = \bar\Sigma^a_{AB}\Phi^a$. 

Introducing the D-instanton, a half of the supersymmetry 
is broken and the unbroken symmetry is generated by 
supercharges $\bar Q_\ad^A$. 
We list these supersymmetry transformations of 
the fields on the D-instanton. 
The fields in the D(-1)-D(-1) sector are 
given by dimensional reduction of the $\cN =4$ super Yang-Mills theory. 
Their transformations are 
\begin{align}
 \delta a_\mu &= 
 \tfrac{i}{2} \bar\xi_{\ad A}\bar\sigma^{\mu\ad\beta}M_\beta^A , \\ 
 \delta\chi^a &= -i\bar\xi_{\ad A} \Sigma^{a AB} \lambda_{\ad B} , \\ 
 \delta M^{\alpha A} &= -\tfrac{1}{2}\bar\xi_{\bd B} \bar\sigma^{mu\bd\alpha} 
 \Sigma^{a AB} [\chi^a,a_\mu] , \\
 \delta\lambda_{\ad A} &= \tfrac12 \bar\xi_{\ad B} \bar\Sigma^{ab\,B}_A 
 [\chi^a,\chi^b] +\bar\xi_{\bd A}\bar\sigma^{\mu\nu\bd}_\ad [a_\mu,a_\nu] . 
\end{align}
By introducing the auxiliary fields $D^c$, 
the transformation of $\lambda_{\ad A}$ becomes 
\begin{equation}
 \delta \lambda_{\ad A} = \tfrac12 \bar\xi_{\bd A} \tau^{c\bd}_\ad D^c . 
\end{equation}
And the transformations in the D3-D(-1) sector are the followings: 
\begin{align}
 \delta\omega_\ad &= -i\bar\xi_{\ad A} \mu^A , \\
 \delta\bar\omega_\ad &= i\bar\xi_{\ad A} \bar\mu^A , \\
 \delta\mu^A &= \varepsilon^{ABCD}\bar\xi_{\ad B} 
 \left(\omega^\ad\chi_{CD} + \Phi_{CD} \omega^\ad\right) , \\
 \delta\bar\mu^A &= \varepsilon^{ABCD}\bar\xi_{\ad B} 
 \left(\chi_{CD}\bar\omega^\ad + \bar\omega^\ad\Phi_{CD}\right) , 
\end{align}
where, $\chi_{AB} = \bar\Sigma^a_{AB}\chi^a$.

\section{Multi-fermion $F$-terms for general $SU(N)$}

In this appendix, we show an explicit expression 
of the superpotential for general $N_f\geq N_c$ 
in the case of unitary gauge group. 
The expression is quite complicated if 
all fields in the D3-D(-1) sector are integrated out. 
Therefore, we do not perform 
all of the integration. 
The superpotential can be expressed in terms 
of the integral with respect to $\omega$: 
\begin{align}
 W = \int d\omega^2d\bar\omega^2\ 
 &\delta^{(3)}(\bar\omega_{\ad u}(\tau^c)^\ad_\bd\omega^{\bd u})\,
 e^{-\frac{1}{2}\bar\omega_{\ad u} 
 \left(Q^u_f Q^{\dag f}_v 
 + \widetilde Q^{\dag u}_f \widetilde Q^f_v\right)\omega^{\ad v}}
 \varepsilon^{u_1\cdots u_{N_c}}\varepsilon_{v_1\cdots v_{N_c}}
 \varepsilon_{f_1\cdots f_{N_f}}\varepsilon^{g_1\cdots g_{N_f}} \notag\\
 &\times\bigg[\frac{1}{N_c!(N_c-2)!(N_f-N_c)!(N_f-N_c+2)!}\notag\\
 &\qquad\times
 Q^{\dagger f_1}_{u_1}\cdots Q^{\dagger f_{N_c-2}}_{u_{N_c-2}} 
 \tQ^{\dagger v_1}_{g_1}\cdots \tQ^{\dagger v_{N_c}}_{g_{N_c}}\notag\\
 &\qquad\times\bar\omega_{\gd u_{N_c-1}}\bar\omega^{\gd}_{u_{N_c}} 
 \bar\psi_{\ad_1 w_1}^{f_{N_c-1}}\omega^{\ad_1 w_1}\cdots
 \bar\psi_{\ad_{N_f-N_c+2} w_{N_f-N_c+2}}^{f_{N_f}}
 \omega^{\ad_{N_f-N_c+2} w_{N_f-N_c+2}}\notag\\
 &\qquad\times\bar\omega_{\bd_1 y_1}\bar{\tp}^{\bd y_1}_{g_{N_c+1}}\cdots 
 \bar\omega_{\bd_{N_f-N_c} y_{N_f-N_c}}
 \bar\tp^{\bd_{N_f-N_c} y_{N_f-N_c}}_{g_{N_f}} \notag\\
 &\quad+ \frac{2}{[(N_c-1)!]^2[(N_f-N_c+1)!]^2}\notag\\
 &\qquad\times Q^{\dagger f_1}_{u_1}\cdots Q^{\dagger f_{N_c-1}}_{u_{N_c-1}} 
 \tQ^{\dagger v_1}_{g_1}\cdots \tQ^{\dagger v_{N_c-1}}_{g_{N_c-1}}\notag\\
 &\qquad\times\bar\omega_{\gd u_{N_c}}\omega^{\gd v_{N_c}}
 \bar\psi_{\ad_1 w_1}^{f_{N_c}}\omega^{\ad_1 w_1}\cdots
 \bar\psi_{\ad_{N_f-N_c+1} w_{N_f-N_c+1}}^{f_{N_f}}
 \omega^{\ad_{N_f-N_c+1} w_{N_f-N_c+1}}\notag\\
 &\qquad\times\bar\omega_{\bd_1 y_1}\bar{\tp}^{\bd y_1}_{g_{N_c}}\cdots 
 \bar\omega_{\bd_{N_f-N_c+1} y_{N_f-N_c+1}}
 \bar\tp^{\bd_{N_f-N_c+1} y_{N_f-N_c+1}}_{g_{N_f}} \notag\\
 &\quad+ \frac{1}{N_c!(N_c-2)!(N_f-N_c)!(N_f-N_c+2)!}\notag\\
 &\qquad\times Q^{\dagger f_1}_{u_1}\cdots Q^{\dagger f_{N_c}}_{u_{N_c}} 
 \tQ^{\dagger v_1}_{g_1}\cdots \tQ^{\dagger v_{N_c-2}}_{g_{N_c-2}}\notag\\
 &\qquad\times\omega_{\gd}^{v_{N_c-1}}\omega^{\gd v_{N_c}} 
 \bar\psi_{\ad_1 w_1}^{f_{N_c+1}}\omega^{\ad_1 w_1}\cdots
 \bar\psi_{\ad_{N_f-N_c} w_{N_f-N_c}}^{f_{N_f}}
 \omega^{\ad_{N_f-N_c} w_{N_f-N_c}}\notag\\
 &\qquad\times\bar\omega_{\bd_1 y_1}\bar{\tp}^{\bd y_1}_{g_{N_c-1}}\cdots 
 \bar\omega_{\bd_{N_f-N_c+2} y_{N_f-N_c+2}}
 \bar\tp^{\bd_{N_f-N_c+2} y_{N_f-N_c+2}}_{g_{N_f}} 
 \bigg].  
\end{align}
The fields on the D-instanton correspond to the moduli of the instanton, 
and $\omega$'s are the gauge direction and the size of the instanton. 
For example, 
the instanton solution of the gauge field $A_\mu$ 
is written in terms of $\omega$ as 
\begin{equation}
 A_\mu(x) = \omega_\ad^u(\bar\sigma_{\mu\nu})^\ad_\bd\bar\omega^\bd_v 
  \frac{(x-x_0)^\nu}
  {(x-x_0)^2[(x-x_0)^2+\frac{1}{2}\bar\omega_{\gd w}\omega^{\gd w}]}, 
\end{equation}
where, we have taken the singular gauge, 
and $x_0^\mu = a^\mu$ is the position of the instanton.

\section{Constrained Gaussian integral}\label{CGauss}
In this appendix, we describe the calculation of 
the constraint Gaussian integral. 
Let us define $F$ by
\begin{equation}
 F=\int d\w\, d\bar{\w}\ \delta^{(3)}(\bar{\w}\tau^c\w)\,
  e^{-\bar{\w}_\ad A \w^\ad}
\end{equation}
where $A$ is an $N\times N$ matrix.
Writing the $\delta$-function in terms of the contour integral, 
we obtain 
\begin{equation}
 F=\int d\w\, d\bar{\w}\,d^3 k\ 
  e^{-\bar{\w}_\ad A \w^\ad+ik_a\bar{\w}_\ad\tau^{a\ad}_{\ \bd}\w^\bd}
  =\int d\Omega\, d\bar{\Omega}\,d^3k\ e^{-\bar{\Omega}\tilde{A}\Omega}
\end{equation}
where $\tilde{A}$ is the $2N \times 2N$ matrix which is defined as 
\begin{equation}
\tilde{A}=\left( \begin{array}{cc}
                  A-ik_3 & -k_1-k_2 \\
                  -ik_1+k_2 & A+ik_3  \end{array}  \right) , 
\end{equation}
and $\Omega$ and $\bar\Omega$ are defined as 
\begin{align}
 \Omega &= 
 \left(
 \begin{matrix}
  \omega^1 \\ \omega^2
 \end{matrix}\right),&
 \bar \Omega &= 
 \left(\begin{matrix}
  \bar\omega_1, & \bar\omega_2
 \end{matrix}\right). 
\end{align}
Then integrating over $\Omega$ and $\bar{\Omega}$, we have
\begin{equation}
F=\int d^3 k \frac{1}{\det \tilde{A}}=\int dk \frac{1}{\det (A^2+k^2)}
\end{equation}
where we used the following formula: 
\begin{equation}
 \det \left( \begin{array}{cc}
                  A & B \\
	     C & D  \end{array}  \right)
 =\det(AD-BC)
\end{equation}
if $A, B, C, D$ mutually commute. 
Diagonalizing $A$ into $diag(a_1,a_2 \cdots a_N)$ and picking up all
residues in the upper half plane of $k$, we obtain 
\begin{equation}
F=\int dk \frac{k^2}{\prod_{m=1}^N(a_m^2+k^2)}
=\sum_{n=1}^{N}\frac{a_n}{\prod_{m\ne n}(a_m^2-a_n^2)}
\end{equation}
This can be written in terms of $A$ as 
\begin{equation}
F=\tr \frac{A}{\Sigma_{r=0}^{N-1} A^{2r}C^{(r)}}
\end{equation}
where 
\begin{equation}
C^{(r)\,i}_{\,\,\,\,j}=\frac{1}{(N-r-1)!r!}
 \varepsilon^{in_1 \cdots n_r \, l_1 \cdots
  l_{N-(r+1)}}\varepsilon_{jn_1\cdots n_r \, k_1 \cdots k_{N-(r+1)}}
(A^2)^{k_1}_{l_1}\cdots (A^2)^{k_{N-(r+1)}}_{l_{N-(r+1)}}
\end{equation}


\begin{thebibliography}{99}

\bibitem{Beasley:2004ys}
  C.~Beasley and E.~Witten,
  ``New instanton effects in supersymmetric QCD,''
  JHEP {\bf 0501} (2005) 056
  [arXiv:hep-th/0409149].

\bibitem{Witten:1995gx}
  E.~Witten,
  ``Small Instantons in String Theory,''
  Nucl.\ Phys.\  B {\bf 460} (1996) 541
  [arXiv:hep-th/9511030].

\bibitem{Douglas:1995bn}
  M.~R.~Douglas,
  ``Branes within branes,''
  arXiv:hep-th/9512077.

\bibitem{Witten:1996bn}
  E.~Witten,
  ``Non-Perturbative Superpotentials In String Theory,''
  Nucl.\ Phys.\  B {\bf 474} (1996) 343
  [arXiv:hep-th/9604030].

\bibitem{Harvey:1999as}
  J.~A.~Harvey and G.~W.~Moore,
  ``Superpotentials and membrane instantons,''
  arXiv:hep-th/9907026.

\bibitem{Blumenhagen:2006xt}
  R.~Blumenhagen, M.~Cvetic and T.~Weigand,
  ``Spacetime instanton corrections in 4D string vacua - the seesaw mechanism
  for D-brane models,''
  Nucl.\ Phys.\  B {\bf 771} (2007) 113
  [arXiv:hep-th/0609191].

\bibitem{Ibanez:2006da}
  L.~E.~Ibanez and A.~M.~Uranga,
  ``Neutrino Majorana masses from string theory instanton effects,''
  JHEP {\bf 0703} (2007) 052
  [arXiv:hep-th/0609213].

\bibitem{Florea:2006si}
  B.~Florea, S.~Kachru, J.~McGreevy and N.~Saulina,
  ``Stringy instantons and quiver gauge theories,''
  JHEP {\bf 0705} (2007) 024
  [arXiv:hep-th/0610003].

\bibitem{Abel:2006yk}
  S.~A.~Abel and M.~D.~Goodsell,
  ``Realistic Yukawa couplings through instantons in intersecting brane
  worlds,''
  JHEP {\bf 0710} (2007) 034
  [arXiv:hep-th/0612110].

\bibitem{Argurio:2007vqa}
  R.~Argurio, M.~Bertolini, G.~Ferretti, A.~Lerda and C.~Petersson,
  ``Stringy Instantons at Orbifold Singularities,''
  JHEP {\bf 0706} (2007) 067
  [arXiv:0704.0262 [hep-th]].

\bibitem{Aharony:2007pr}
  O.~Aharony and S.~Kachru,
  ``Stringy Instantons and Cascading Quivers,''
  JHEP {\bf 0709} (2007) 060
  [arXiv:0707.3126 [hep-th]].

\bibitem{Billo:2002hm}
  M.~Billo, M.~Frau, I.~Pesando, F.~Fucito, A.~Lerda and A.~Liccardo,
  ``Classical gauge instantons from open strings,''
  JHEP {\bf 0302} (2003) 045
  [arXiv:hep-th/0211250].

\bibitem{Bertolini:2001gg}
  M.~Bertolini, P.~Di Vecchia, G.~Ferretti and R.~Marotta,
  ``Fractional branes and N = 1 gauge theories,''
  Nucl.\ Phys.\  B {\bf 630} (2002) 222
  [arXiv:hep-th/0112187].

\bibitem{Affleck:1983mk}
  I.~Affleck, M.~Dine and N.~Seiberg,
  ``Dynamical Supersymmetry Breaking In Supersymmetric QCD,''
  Nucl.\ Phys.\  B {\bf 241} (1984) 493.

\bibitem{Intriligator:1995au}
  K.~A.~Intriligator and N.~Seiberg,
  ``Lectures on supersymmetric gauge theories and electric-magnetic  duality,''
  Nucl.\ Phys.\ Proc.\ Suppl.\  {\bf 45BC} (1996) 1
  [arXiv:hep-th/9509066].

\bibitem{Akerblom:2006hx}
  N.~Akerblom, R.~Blumenhagen, D.~Lust, E.~Plauschinn and M.~Schmidt-Sommerfeld,
  ``Non-perturbative SQCD Superpotentials from String Instantons,''
  JHEP {\bf 0704} (2007) 076
  [arXiv:hep-th/0612132].

\bibitem{Seiberg:1994bz}
  N.~Seiberg,
  ``Exact Results On The Space Of Vacua Of Four-Dimensional Susy Gauge
  Theories,''
  Phys.\ Rev.\  D {\bf 49} (1994) 6857
  [arXiv:hep-th/9402044].

\bibitem{Atiyah:1978ri}
  M.~F.~Atiyah, N.~J.~Hitchin, V.~G.~Drinfeld and Yu.~I.~Manin,
  ``Construction of instantons,''
  Phys.\ Lett.\  A {\bf 65} (1978) 185.

\bibitem{Dorey:2002ik}
  N.~Dorey, T.~J.~Hollowood, V.~V.~Khoze and M.~P.~Mattis,
  ``The calculus of many instantons,''
  Phys.\ Rept.\  {\bf 371} (2002) 231
  [arXiv:hep-th/0206063].

\bibitem{Witten}
 E. Witten, ``Worldsheet corrections via D-instantons,''
 JHEP {\bf 0001} (2000) 030
  [arXiv:hep-th/9907041].

\bibitem{Blumenhagen:2007bn}
  R.~Blumenhagen, M.~Cvetic, R.~Richter and T.~Weigand,
  ``Lifting D-Instanton Zero Modes by Recombination and Background Fluxes,''
  JHEP {\bf 0710} (2007) 098
  [arXiv:0708.0403 [hep-th]].

\bibitem{Douglas:1996sw}
  M.~R.~Douglas and G.~W.~Moore,
  ``D-branes, Quivers, and ALE Instantons,''
  arXiv:hep-th/9603167.

\bibitem{Gimon:1996rq}
  E.~G.~Gimon and J.~Polchinski,
  ``Consistency Conditions for Orientifolds and D-Manifolds,''
  Phys.\ Rev.\  D {\bf 54} (1996) 1667
  [arXiv:hep-th/9601038].

\bibitem{Intriligator:1995ne}
  K.~A.~Intriligator and P.~Pouliot,
  ``Exact superpotentials, quantum vacua and duality in supersymmetric SP(N(c))
  gauge theories,''
  Phys.\ Lett.\  B {\bf 353} (1995) 471
  [arXiv:hep-th/9505006].

\bibitem{Intriligator:1995id}
  K.~A.~Intriligator and N.~Seiberg,
  ``Duality, monopoles, dyons, confinement and oblique confinement in
  supersymmetric SO(N(c)) gauge theories,''
  Nucl.\ Phys.\  B {\bf 444} (1995) 125
  [arXiv:hep-th/9503179].




\end{thebibliography}
\end{document}